\documentclass{sig-alternate-10pt}
\makeatletter

\newcommand{\revise}[1]{{\color{black} #1}}
\newcommand{\zhixiong}[1]{{\color{black} #1}}

\usepackage{graphicx,cite}
\usepackage{float}
\usepackage[font=small,labelfont=bf]{caption}
\usepackage{subcaption}
\usepackage{todonotes}

\usepackage[shortlabels]{enumitem}
\setitemize{itemsep=3pt,topsep=1pt,parsep=1pt,partopsep=1pt}
\usepackage{times}
\usepackage{microtype}
\newcommand{\subparagraph}{}
\usepackage[compact]{titlesec}
\usepackage{array}

\usepackage{balance}
\usepackage{lipsum}

\usepackage[hyphens]{url}
\usepackage{hyperref}
\usepackage{cleveref}
\crefname{section}{\S}{\S\S}
\Crefname{section}{\S}{\S\S}
\crefformat{section}{\S#2#1#3}
\hypersetup{colorlinks={true},linkcolor={blue},citecolor=blue}

\usepackage{multirow}

\begin{document}
\graphicspath{{figures/}}

\title{Benchmarking NFV Software Dataplanes}


\numberofauthors{1}
\author{
\alignauthor
Zhixiong Niu, Hong Xu, Yongqiang Tian, Libin Liu, Peng Wang, Zhenhua Li$^*$ \\
NetX Lab, City University of Hong Kong\\
$^*$School of Software, Tsinghua University
}

\maketitle

\begin{abstract}

A key enabling technology of NFV is software dataplane, which has attracted much attention in both academia and industry recently. 
Yet, till now there is little understanding about its performance in practice.
In this paper, we make a benchmark measurement study of NFV software dataplanes in terms of 
packet processing capability, one of the most fundamental and critical performance metrics.
Specifically, we compare two state-of-the-art open-source NFV dataplanes, SoftNIC and ClickOS, using commodity 10GbE NICs under various typical workloads.
Our key observations are that (1) both dataplanes have performance issues processing small ($\le$128B) packets; (2) it is not always best to put all VMs of a service chain on one server due to NUMA effect.
We propose resource allocation strategies to remedy the problems, including carefully adding CPU cores and vNICs to VMs, and spreading VMs of a service chain to separate servers.  
To fundamentally address these problems and scale their performance, SoftNIC and ClickOS could improve the support for NIC queues and multiple cores. 



\end{abstract}


\section{Introduction}
\label{sec:intro}

Middleboxes are ubiquitous in today's networks and provide important network functions to operators \cite{SHSK12}.
Traditionally, middleboxes are deployed as dedicated proprietary hardware. Network function virtualization (NFV) is now emerging to replace hardware boxes with virtual software instances running on commodity servers. NFV holds great promises to improve flexibility and efficiency of network function management. Thus it is quickly gaining momentum in the industry \cite{ET12}.
The ETSI Industry Specification Group for NFV has attracted over 290 individual companies as members, including major service providers such as AT\&T, NTT, Sprint, China Mobile, China Unicom, and IT vendors such as Cisco, Huawei, IBM, and VMware \cite{etsi-nfv}.
The NFV market is projected to grow more than 5-fold through 2019 to reach \$11.6 billion \cite{nfvmarket}.

A key enabling technology to NFV is software dataplane. It provides a virtualized platform for hosting software middleboxes with high performance packet I/O to userspace.
A number of NFV dataplanes have been developed recently.
For example SoftNIC \cite{HJPP15} and NetVM \cite{HRW14} use KVM for virtualization and Intel DPDK for packet I/O. ClickOS \cite{MARO14} relies on Xen and netmap \cite{R12} instead for virtualization and packet I/O, respectively.

Despite the progress, there is a lack of understanding on NFV dataplane performance in the community.
\revise{A software dataplane desires high-performance packet processing, flexible programming interfaces, security/isolation between colocating VNFs, and so forth~\cite{MARO14,HRW14}. Among these performance metrics, packet processing capability is fundamental and critical since it determines the basic usability of a software dataplane. Hence it becomes the focus of our study.}
Specifically, we ask the question, {\em how well do these software dataplanes perform packet processing in practice?}

Existing work and their evaluation do not address this question well. Most work (e.g. SoftNIC \cite{HJPP15}) focuses on raw packet I/O without software middleboxes running as VMs on top. Some (e.g. ClickOS \cite{MARO14}) report performance of different software middleboxes only when deployed individually. More importantly, no performance comparison is done across NFV dataplanes under the same environment. Thus, it is unclear whether these NFV dataplanes can achieve line rate with different packet processing logic in software middleboxes, what are their bottlenecks if any, and how they perform against each other in various deployment settings such as NF chaining and colocation?

In this paper, we present arguably the first measurement study of NFV dataplanes that provides initial answers to the above question. We strategically choose two popular open source NFV dataplanes, SoftNIC \cite{HJPP15} and ClickOS \cite{MARO14}, that differ widely in virtualization and packet I/O technologies. As a first step we focus on their packet processing throughput running on commodity servers and 10GbE NICs. We use two basic virtual network functions (VNFs): L3 forwarding and firewall.


Our measurements reveal several major findings:
\begin{enumerate}

\item Both SoftNIC and ClickOS can achieve line rate with medium to large packets ($>$128B), even when CPU is clocked down to 1.2GHz. In a practical setting with mixed packet sizes and low network utilization, both dataplanes can handle typical traffic.
\item Both dataplanes cannot achieve line rate processing small packets ($\le$ 128B) on a 2.6GHz CPU. For SoftNIC the bottleneck is due to the lack of multi-queue support at vNIC. We observe that adding more vNICs and correspondingly more vCPUs achieves line rate for 64B packets. For ClickOS we believe the bottleneck is its high CPU usage, which cannot be resolved due to its lack of SMP support. 
\item Performance also degrades in the NF chaining scenario, with ClickOS being more sensitive to chain length. \revise{Perhaps surprisingly, placing all VNFs of the chain on the same server does not necessarily lead to best performance, because the NUMA effect may further degrade performance when there are too many VNFs to be put to the same CPU socket.  
In this case simply assigning VNFs to different servers and using NICs to chain them can eliminate the NUMA effect.}
\end{enumerate}

The results provide useful implications for efficient resource management of NFV deployment in practice.
For a telecom or ISP that deploys NFV to run her middleboxes, our results suggest that a dynamic resource allocation strategy can be adopted to opportunistically adjust the CPU speed or number of cores of the VNF and save energy without sacrificing performance.
\revise{Most production networks are mildly utilized, suggesting that significant savings of electricity cost can be realized using this approach. Our results on NF chaining also shed light on VNF placement, an important management task of an NFV cluster.}
\revise{We show that it is better to place VNFs on separate servers (on the same CPU socket) and chain them up using NICs in order to eliminate the NUMA effect, when it is impossible to assign them to one CPU socket.}

Our study also provides helpful implications for the research community on performance optimization of software dataplane. The results consistently suggest that an important research direction is the support for multiple cores and NIC queues, which can fundamentally scale the performance of software dataplane in demanding scenarios, especially as the network evolves to 40G and beyond.


\section{Background}
\label{sec:background}

We start by providing background of SoftNIC and ClickOS.

\subsection{SoftNIC}
\label{sec:softnic}
The SoftNIC software dataplane composes of three components: Intel DPDK \cite{dpdk} as the high-performance userspace packet I/O framework, SoftNIC \cite{HJPP15} as the programmable dataplane, and KVM as the hypervisor to isolate the VNFs. 

\noindent{\bf DPDK.} 
The Intel DPDK framework allows applications to poll data directly from the NIC without kernel involvement, thus providing high-performance userspace network I/O. To achieve line rate, a DPDK process occupies the CPU core and constantly polls the NIC for packets. 

\noindent{\bf SoftNIC.}
SoftNIC \cite{HJPP15} is a programmable dataplane abstraction layer that allows developers to flexibly build software that leverages NIC features with minimal performance loss. 
One can develop her own packet processing pipeline with a series of {\em modules}. 
A module can interact with a physical NIC (pNIC) and/or a vNIC of a VM.
When two modules are connected in a pipeline, a {\em traffic class} (TC) is created. A TC is assigned with a unique {\em worker} thread running on a dedicated core to move packets between the modules. 
A worker may be assigned to multiple TCs. 

The open source version of SoftNIC has recently been renamed BESS \cite{bess}. We do not use the name here to avoid confusion. 

\noindent{\bf KVM.} 
SoftNIC provides a backend vNIC driver based on {\tt vhost-net} \cite{vhost-net} which allows it to interact with KVM. We thus choose KVM as the hypervisor environment for it.

\subsection{ClickOS}
\label{sec:clickos}

\noindent{\bf ClickOS.} ClickOS \cite{MARO14} is another popular NFV platform.  
It composes of netmap \cite{R12} and VALE \cite{rizzo2012vale} as the packet I/O framework, Click \cite{K01} as the programmable dataplane, and Xen as the hypervisor. 
By redesigning the virtual network drivers in Xen, ClickOS achieves very high packet processing performance. Meanwhile, by leveraging Click users can flexibly build software middleboxes. 

\noindent{\bf VALE and netmap.} VALE is a virtual software switch for packet I/O between VMs based on netmap \cite{R12}. ClickOS modifies VALE to support pNIC directly. A key difference between VALE and TCs in SoftNIC (or any DPDK based software dataplane) is that VALE does not use dedicated threads/cores to move packets between modules; the sending thread does the work of copying packets into the Rx queue.


\section{Methodology}
\label{sec:method}
We explain our measurement methodology in detail here.

\subsection{Hardware Setup}
\label{sec:hardware}
We conduct our measurements using {\em physical} machines rent from Aptlab\cite{aptlab}. We use two c6220 nodes with 2 Xeon E5-2650v2 processors (8 cores each, 2.6Ghz), 64GB DDR3 1.86GHz Memory and an Intel X520 10GbE PCIe dual port NIC. For most of the experiments, one node runs a packet generator to send packets of different sizes to the other node, which serves as the hypervisor hosting VNFs to process packets. Packets are sent back through another NIC of the hypervisor to the first node, which is also our vantage point. 
We disable DFVS and fix the CPU at 2.6Ghz for both nodes unless otherwise stated. For NICs, we disable auto-negotiation, TSO, and GSO as recommended by ClickOS\cite{MARO14}.

\subsection{Software Dataplane Settings}
\label{sec:software}

We met some difficulties in deploying SoftNIC and ClickOS on our testbed. Since their components are independently maintained and some have evolved, we were unable to build them in the same environment reported in the original papers or the available online documentation. Some components, such as netmap, VALE, and Xen, have strict dependencies on the kernel version, NIC models, and hardware features which further complicate the problem.
With the help of developers for both dataplanes, we experimented with over a dozen different environments, and found settings that yield the best performance.\footnote{For ClickOS, we found only one environment where we can successfully build all the components as in \cite{MARO14} in our testbed.} 
For SoftNIC, we use Linux kernel 3.19.0, QEMU/KVM 2.2.1, DPDK 2.20, and the latest SoftNIC source code \cite{bess}. 
For ClickOS, we use Linux kernel 3.9.10, Xen 4.4.2\cite{BDFH03}, ClickOS 0.1 \cite{clickos-code} and netmap commit {\tt 3ccdada} \cite{netmap-code} respectively. We use an older version of netmap that works with the modified Xennet library \cite{xennet} provided by ClickOS for Xen backend and frontend NIC drivers.

We confirmed our settings with the authors of SoftNIC and ClickOS to ensure validity of our measurements. We also verified that the baseline performance of L2 forwarding is consistent with or no worse than those reported previously in \cite{HJPP15} and \cite{MARO14}.

\begin{figure}[ht]
    \captionsetup{justification=centering}
    \centering
    \includegraphics[width=.85\linewidth]{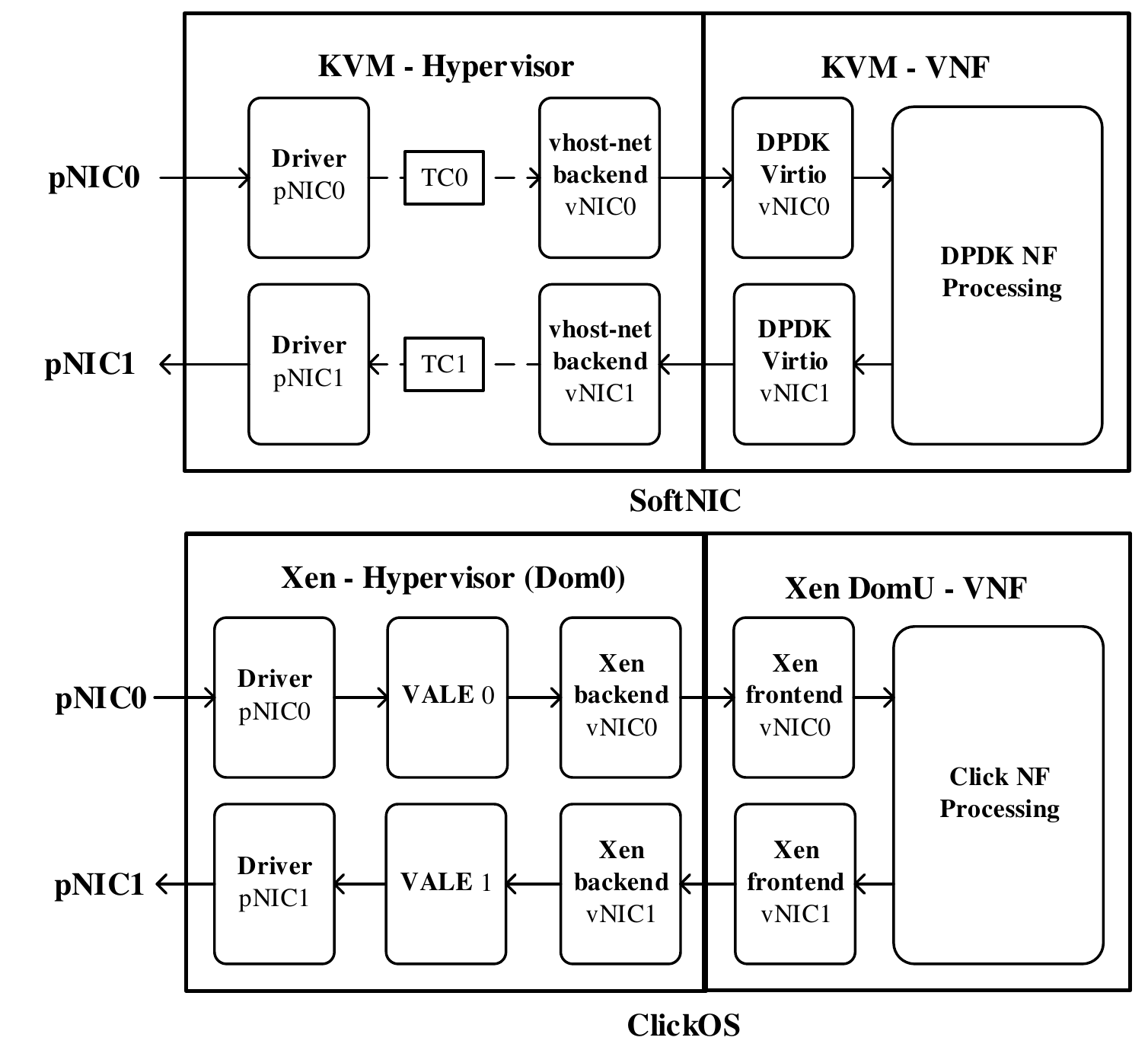}
    \caption{Packet I/O pipeline with a single VNF in our measurement.}
    \label{fig:topology}
    \vspace{-3mm}
\end{figure}

Figure~\ref{fig:topology} illustrates the packet I/O pipeline we use in our measurements with a single VNF. 
Each VNF has two vNICs, vNIC0 as the ingress NIC and vNIC1 as the egress NIC to its next hop. 
In general, for both SoftNIC and ClickOS, first a packet is moved by the pNIC0 driver to the backend of vNIC0, which then sends it to the frontend driver in the VNF. After being processed by the VNF, the packet is sent to the frontend and then backend of vNIC1. Finally it is sent to an output NIC in the hypervisor.
The difference between the two software dataplanes is that, SoftNIC uses a TC to connect a pNIC and a VNF (or two VNFs), while ClickOS uses a VALE switch instead. NF chaining can be realized by having the TC or VALE connecting the vNIC1 backend driver of the previous VNF to the vNIC0 backend of the next VNF in the chain.

For maximum performance, we pin vCPU(s) of VNFs into fixed physical cores by {\tt taskset} in KVM and {\tt xl vcpu-pin} in Xen. As mentioned in \cref{sec:softnic} TCs in SoftNIC need to be assigned with workers which also must be pinned into dedicated cores. We pin vCPUs to cores in the same socket whenever possible to avoid severe performance penalty caused by NUMA \cite{levinthal2009performance} (more in \cref{sec:result_chain}).

\subsection{Network Functions}
\label{sec:NF_used}
We use two NFs in this study, L3 forwarding (L3FWD), and firewall.
We cannot use generic software such as Snort or Bro because they are not written with DPDK or Click to exploit software dataplane for packet processing. 
We assign 1 vCPU and 1GB memory to each VNF unless stated otherwise. 


\noindent{\bf L3FWD.} 
We use the {\tt ip\_pipeline} example code provided by DPDK as the L3FWD implementation in SoftNIC.
The ClickOS implementation is done by concatenating \texttt{FromDevice}, \texttt{StaticIPLookup}, and \texttt{ToDevice} elements. In both implementations we insert 10 entries to the routing table. 


\noindent{\bf Firewall.}
We build the firewall implementation based on the same {\tt ip\_pipeline} in SoftNIC. 
In ClickOS, the implementation uses \texttt{FromDevice}, \texttt{IPFilter}, and \texttt{ToDevice} elements. 
We use 10 rules to filter packets.

We also use the simple L2 forwarding in the NF chaining experiment only though. We do not consider it a NF as it does not have any packet processing logic.

\noindent{\bf L2FWD.} 
We directly use the L2FWD provided out-of-the-box from {\tt ip\_pipeline} of DPDK for SoftNIC. In ClickOS, we implement L2FWD by connecting the \texttt{FromDevice} and \texttt{ToDevice} elements between two vNICs. 


\subsection{Miscellaneous }

We use the DPDK {\tt pkt-gen} module to generate packets for experiments with SoftNIC, and the netmap {\tt pkt-gen} for ClickOS. We use different packet sizes: 64B, 128B, 256B, 512B, 1024B, and 1500B. 
\revise{We also use an empirical packet size distribution from Facebook's data center network \cite{RZBP15} to see how the dataplanes perform in a practical environment}. In all scenarios, we verify that the {\tt pkt-gen}s can achieve line rate of 10Gbps.
We mainly use throughput in both million packets per second (Mpps) and Gbps as the performance metrics.


\section{Results}
\label{sec:result}
We investigate the performance of SoftNIC and ClickOS in different scenarios in this section. 
The thesis of the evaluation is simple: can these NFV dataplanes achieve line rate, and if not, what are the bottlenecks?
We first look at the baseline scenario with a single software middlebox, running different NFs with varying CPU speed (\cref{sec:frequency}). Based on the results we analyze and identify performance bottlenecks of both dataplanes (\cref{sec:cores}). We then deploy multiple NFs in two scenarios that are commonplace in practice: NF chaining where packets go through the middleboxes sequentially for processing (\cref{sec:result_chain}), and NF colocation where multiple NFs colocate on the same server and work independently (\cref{sec:colocation}).

\subsection{Baseline Performance}
\label{sec:frequency}

\begin{figure*}[!th]
    \vspace{-1mm}
    \captionsetup{justification=centering}
    \centering
    \includegraphics[width=1.0\textwidth]{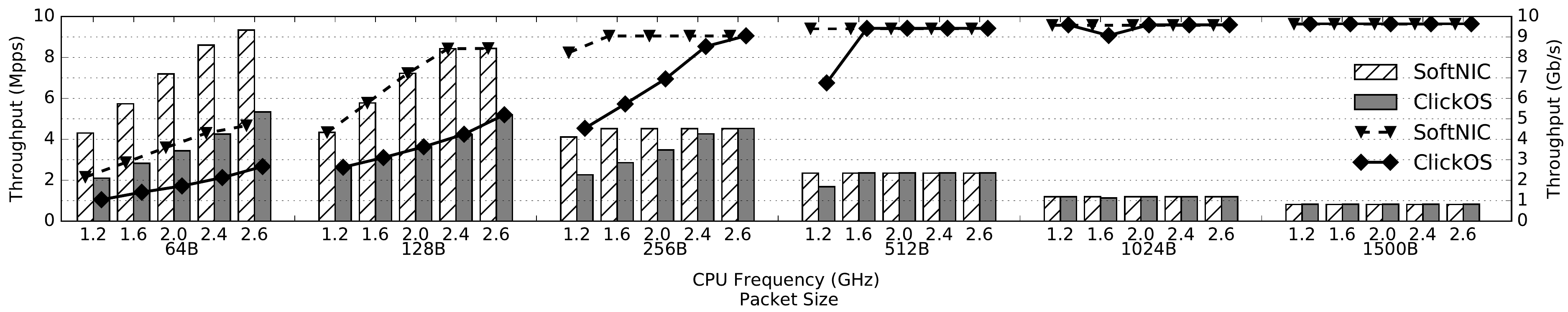}
    \caption{Throughput with different CPU speeds and packet sizes. We show throughput in both Mpps and Gb/s.}
    \label{fig:CPU_freq}
    \vspace{-5mm}
\end{figure*}

We start with just a single VNF. Since software packet processing is CPU-intensive, we want to see if CPU speed is the bottleneck here. In this set of experiments, we vary the CPU speed from the configurable range of 1.2GHz to 2.6GHz for our CPU without Turbo Boost, and investigate the throughput with different packet sizes. We modify the CPU frequency using {\tt cpufreq-set} and {\tt xenpm set-scaling-speed} for SoftNIC and ClickOS, respectively.


Figure~\ref{fig:CPU_freq} demonstrates the performance of L3FWD. We observe the following. 
First, performance increases with CPU speed for small packets (64B--256B), which is expected---a faster CPU can process more instructions and thus more packets. For 64B and 128B packets, performance improvement is commensurate with CPU speed-up between 1.2GHz to 2.4GHz. With 64B packet for instance, at 1.2GHz throughput of SoftNIC and ClickOS is 4.31Mpps and 2.10Mpps, respectively, while at 2.4GHz it roughly doubles at 8.60Mpps and 4.26Mpps, respectively. The improvement is smaller at 2.6GHz.
Second, both NFV dataplanes achieve line rate for packets bigger than 128B, but have problems dealing with smaller packets even at 2.6GHz.
SoftNIC can achieve 10Gbps with 128B packets at 2.4GHz, and ClickOS can process 256B packets at 10Gbps at 2.6GHz. Yet for 64B packets, even at 2.6GHz, neither achieves line rate: SoftNIC tops at 9.34Mpps and ClickOS 5.34Mpps. 
Third, SoftNIC outperforms ClickOS in all cases, especially for small packets. We compare our results with those reported in the ClickOS original paper \cite{MARO14} and confirm they are similar. For example, in Figure~13 of \cite{MARO14}, throughput of L3FWD and Firewall with 64B packets are 4.25Mpps and 5.40Mpps, respectively, and ours are 5.34Mpps and 5.03Mpps, respectively.
Finally, we find that the performance difference between L3FWD and firewall is minimal, as shown in Table~\ref{table:rst_3NF}. We thus only show L3FWD results hereafter for brevity. 

\begin{table}[ht]
\vspace{-1mm}
\centering
\caption{Throughput of different NFs with varying CPU speed. Packet size is 64 bytes. }
\label{table:rst_3NF}
\small
\begin{tabular}{|l|l|l|l|l|}
\hline
          & \multicolumn{2}{l|}{SoftNIC (Mpps)} & \multicolumn{2}{l|}{ClickOS (Mpps)} \\ \hline
CPU     & L3FWD     & Firewall      & L3FWD     & Firewall    \\ \hline
1.2GHz       & 4.31     & 4.45            & 2.10    & 1.98       \\ \hline
1.6GHz      & 5.73    & 5.92           & 2.82     & 2.75       \\ \hline
2.0GHz        & 7.19     & 7.38          & 3.44     & 3.27       \\ \hline
2.4GHz       & 8.60     & 8.87            & 4.26     & 4.24       \\ \hline
2.6GHz      & 9.34     & 9.59           & 5.34     & 5.03       \\ \hline
\end{tabular}
\vspace{-2mm}
\end{table}

\zhixiong{We also use an empirical packet size distribution from Facebook's web server cluster \cite{RZBP15} to see how the software dataplanes perform in a practical environment. The median packet size is $\sim$120B, and most packets are less than 256B. We configure {\tt pkt-gen} to sample the trace and generate packets first at 10Gbps, and observe that the average throughput in SoftNIC is 8.662Mpps. However a production network is rarely fully utilized. Facebook reports their median link utilization is 10\%--20\% \cite{RZBP15}. 
This implies that the median packet processing requirement is 0.87Mpps--1.73Mpps. Both SoftNIC and ClickOS are able to provide such capability in lowest CPU frequency of 1.2GHz. }

\revise{These observations have interesting implications to NFV resource allocation. They suggest that there are ample opportunities for the operator to downclock the CPU in order to save energy and electricity cost in the average case. Care has to be taken though, of course, to ensure performance does not suffer when there are sudden bursts of small packets.
This may be a useful resource allocation strategy for operators as well as meaningful research directions to look into for the networking community. Our observations also motivate us to identify performance bottlenecks in both NFV dataplanes for small packets, which we explain next. }



\subsection{Performance Bottlenecks}
\label{sec:cores}

One may argue that the performance deficiency of software dataplanes in small packet regime is acceptable in practice, since small packets may be less common. However, these systems may not be able to achieve line rates in the emerging 40G or 100G networks \cite{JYAV15}, even for large packets. Therefore we believe it is important for us to understand the performance bottleneck and improve performance. 

\begin{figure}[th!]
    \centering
    \includegraphics[width=0.8\linewidth]{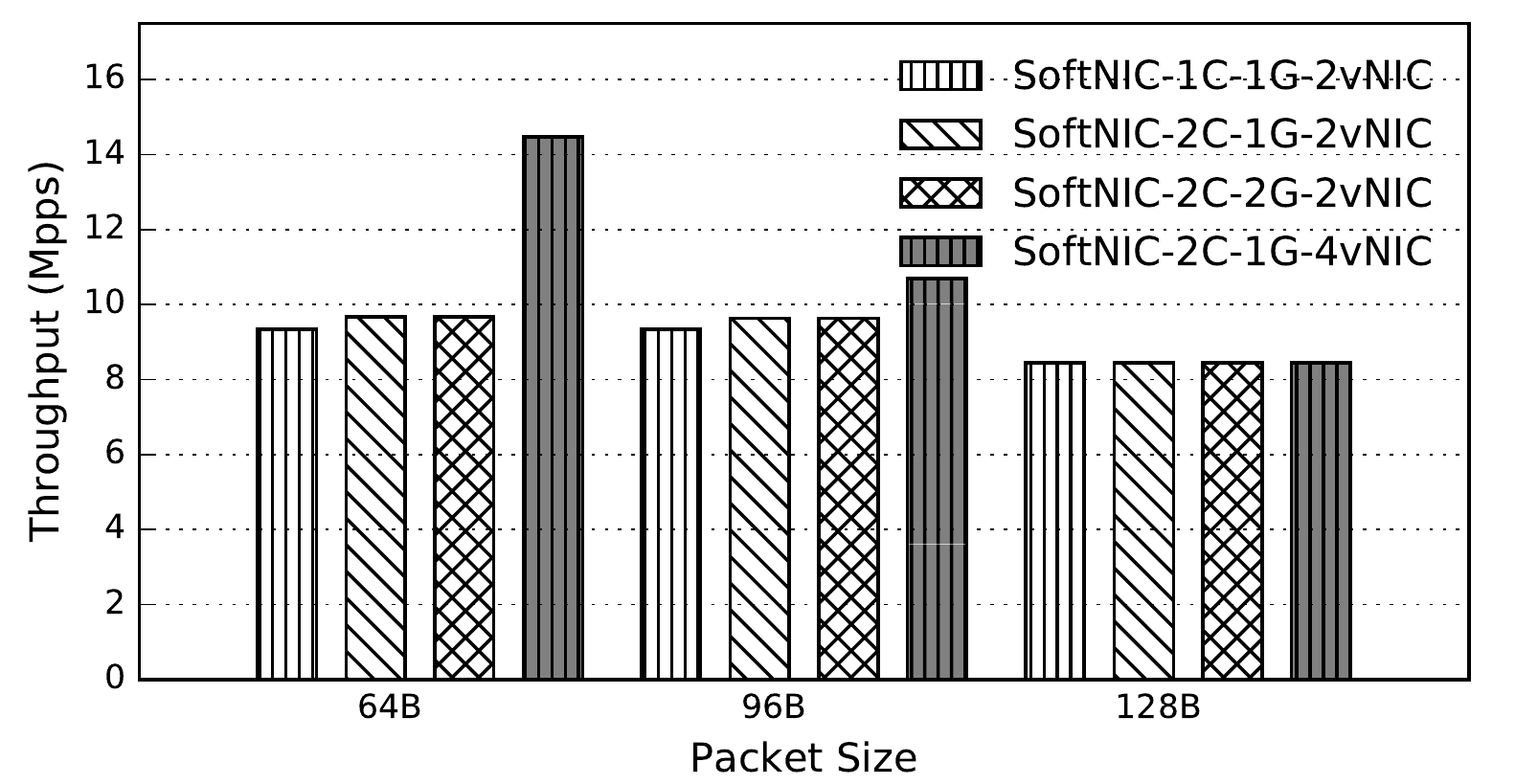}
     \caption{Small packet throughput of SoftNIC with different resource allocations. For example ``SoftNIC-1C-1G-2vNICs'' means we use 1 vCPU, 1GB RAM, and 2vNICs (one for input and one for output) for the VM. For the last setting, we use 2 vCPUs, 1GB RAM, and 4vNICs (two for input and two for output). }
    \label{fig:bottleneck}
\vspace{-3mm}
\end{figure}

\begin{figure}[ht]
\vspace{-2mm}
    \centering
    \includegraphics[width=0.8\linewidth]{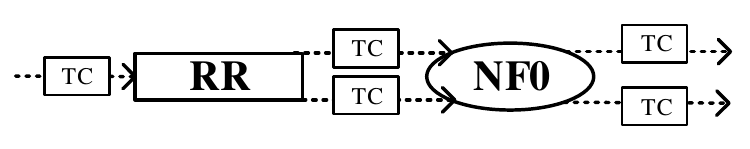}
    \caption{The pipeline of 2 NFs with a round-robin module in SoftNIC. }
    \label{fig:rrtop}
    \vspace{-2mm}
\end{figure}

To identify bottlenecks, we conduct the following analysis. 
For SoftNIC, we observe from using {\tt monitor port} command that about 5Mpps 64B packets are lost in the pipeline between pNIC0 and vNIC0. 
To see if VMs are the bottleneck, we allocate more vCPUs and memory and observe the L3FWD throughput with 64B, 96B, and 128B packets. As shown in Figure~\ref{fig:bottleneck}, however, this results in little improvement. After discussing with SoftNIC authors, we suspect that the bottleneck is the {\tt vhost-net} queue in vNIC0. SoftNIC currently does not support multiple {\tt vhost-net} queues for vNIC, which explains why adding resources to the VM does not help. 
To verify the analysis, we conduct another experiment by adding a round-robin module (RR) and another two vNICs to the L3FWD VM as shown in Figure~\ref{fig:rrtop}. Traffic is evenly split between the two input vNICs. 
This time throughput reaches line rates for all packet sizes as shown in Figure~\ref{fig:bottleneck}. 

For {ClickOS}, we analyze the CPU utilization of the L3FWD instance with the CPU at the highest 2.6GHz. As shown in Table~\ref{cpu}, ClickOS uses 100\% CPU when processing 64B and 128B packets, and larger packets lead to much lower CPU utilization. This implies that more CPU resource may be needed here. However, ClickOS currently does not have SMP support \cite{MARO14}, preventing us from adding more cores to the VM. This also means adding more vNICs does not help without more CPU. Another possible solution is to use multiple VNFs working in parallel. This naturally requires a load balancer (LB) to split the traffic. VALE is a simple L2 switch without any load balancing capability \cite{rizzo2012vale}. Adding a LB VM does not work either since the LB itself becomes the bottleneck.
Therefore we are unable to resolve the bottleneck without modification to ClickOS itself. 

\begin{table}[h]
\centering
\small
\caption{The CPU utilization of L3FWD in ClickOS. CPU is at 2.6GHz. }
\label{cpu}
\begin{tabular}{llllll}
\hline
\multicolumn{1}{|l|}{64B}   & \multicolumn{1}{l|}{128B}  & \multicolumn{1}{l|}{256B}   & \multicolumn{1}{l|}{512B} & \multicolumn{1}{l|}{1024B}  & \multicolumn{1}{l|}{1500B}  \\ \hline
\multicolumn{1}{|l|}{100\%} & \multicolumn{1}{l|}{100\%} & \multicolumn{1}{l|}{70.1\%} & \multicolumn{1}{l|}{63\%} & \multicolumn{1}{l|}{53.6\%} & \multicolumn{1}{l|}{53.8\%} \\ \hline
                            &                            &                             &                           &                             &                          
\end{tabular}
\vspace{-6mm}
\end{table}

To summarize, the results here verify that SoftNIC's bottleneck is the {\tt vhost-net} queue of the vNIC. This can be resolved by sending traffic to two vNICs of one VNF in parallel to fully utilize multiple vCPUs. A complete fix requires SoftNIC to add support for multiple {\tt vhost-net} queues. We have confirmed our analysis with the SoftNIC team already. We also present evidence to suggest that ClickOS should add SMP support that allows it to utilize multiple CPU cores. In any case, we note that it is imperative for the NFV software dataplane architecture to provide horizontal scaling of its performance, in order to better utilize multiple cores and physical NIC queues. We believe this is an interesting open research area as the NICs evolves to 40Gbps and beyond.


\subsection{NF Chaining}
\label{sec:result_chain}

It is common to deploy multiple software middleboxes on the same machine. In this section we look into the NF chaining scenario, where the processing pipeline consists of a chain of different middleboxes. We are interested to see if the performance of an NF chain can match that of just a single NF. 
We compose chains of different lengths: the 1-NF chain uses only a L3FWD; the 2-NF chain uses a firewall followed by a L3FWD; and the 3-NF chain adds a L2FWD to the end of the 2-NF chain. 

\begin{figure}[ht]
    \centering
    \vspace{-0.5mm}
    \includegraphics[width=0.9\linewidth]{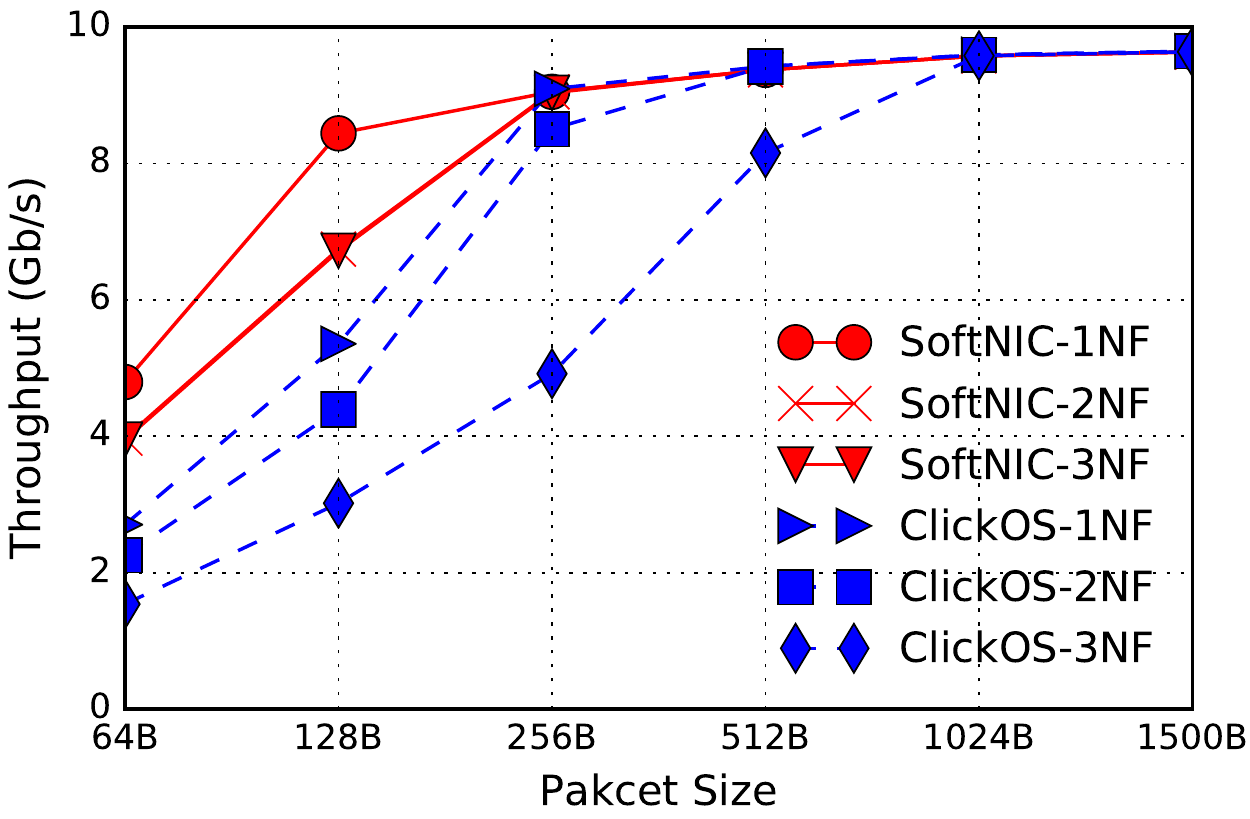}
    \caption{Throughput with varying length of NF chain.}
    \label{fig:rst_chain}
    \vspace{-2mm}
\end{figure}

The results are shown in Figure~\ref{fig:rst_chain} with all vCPUs running on the same physical CPU. The performance of SoftNIC suffers mild degradation for 64B--128B packets, as can be seen from the overlapping lines of 2-NF chain and 3-NF chain. We suspect there is a bottleneck in chaining the vNICs of different VMs because both firewall and L3FWD can achieve higher performance individually as shown in \cref{sec:frequency}.  
Performance of ClickOS also degrades as the chain grows, especially for small packets. 
A ClickOS L3FWD achieves line rate with 256B packets, but a 2-NF chain or 3-NF chain cannot. A 3-NF chain cannot even reach line rate with 512B packets. 
Note that here we use multiple VALE switches with independent vCPUs pinning to different cores to chain the VMs as suggested by \cite{MARO14}. We believe the overhead of copying packets in VALE attributes to the performance penalty.

\begin{figure}[ht]
\vspace{-2mm}
    \centering
    \includegraphics[width=0.9\linewidth]{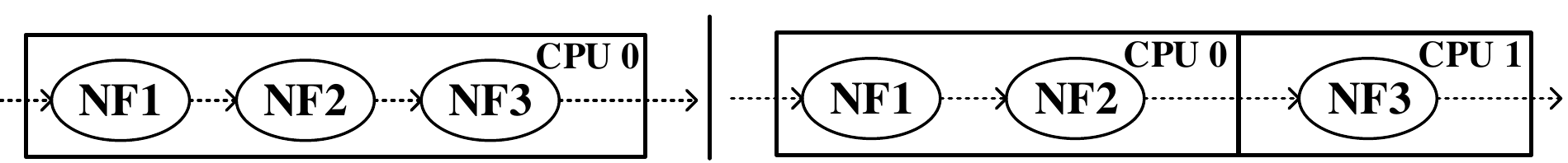}
    \caption{Pipeline of NF-Chains}
\vspace{-4mm}
    \label{fig:top_chain}
\end{figure}

When deploying a NF chain, an important factor we must consider is the affinity of vCPUs and the effect of NUMA. 
As an example Figure~\ref{fig:top_chain} depicts two possibilities of vCPU settings with a 3-NF chain. We can pin each vCPU to the same physical CPU, or pin them to CPUs in different sockets. The latter is unavoidable sometimes as the commodity CPUs have limited cores per CPU, and DPDK-based NFV dataplanes like SoftNIC require many dedicated cores as mentioned in \cref{sec:softnic} and \cref{sec:software}.

\begin{figure}[ht]
    \centering
    \includegraphics[width=0.9\linewidth]{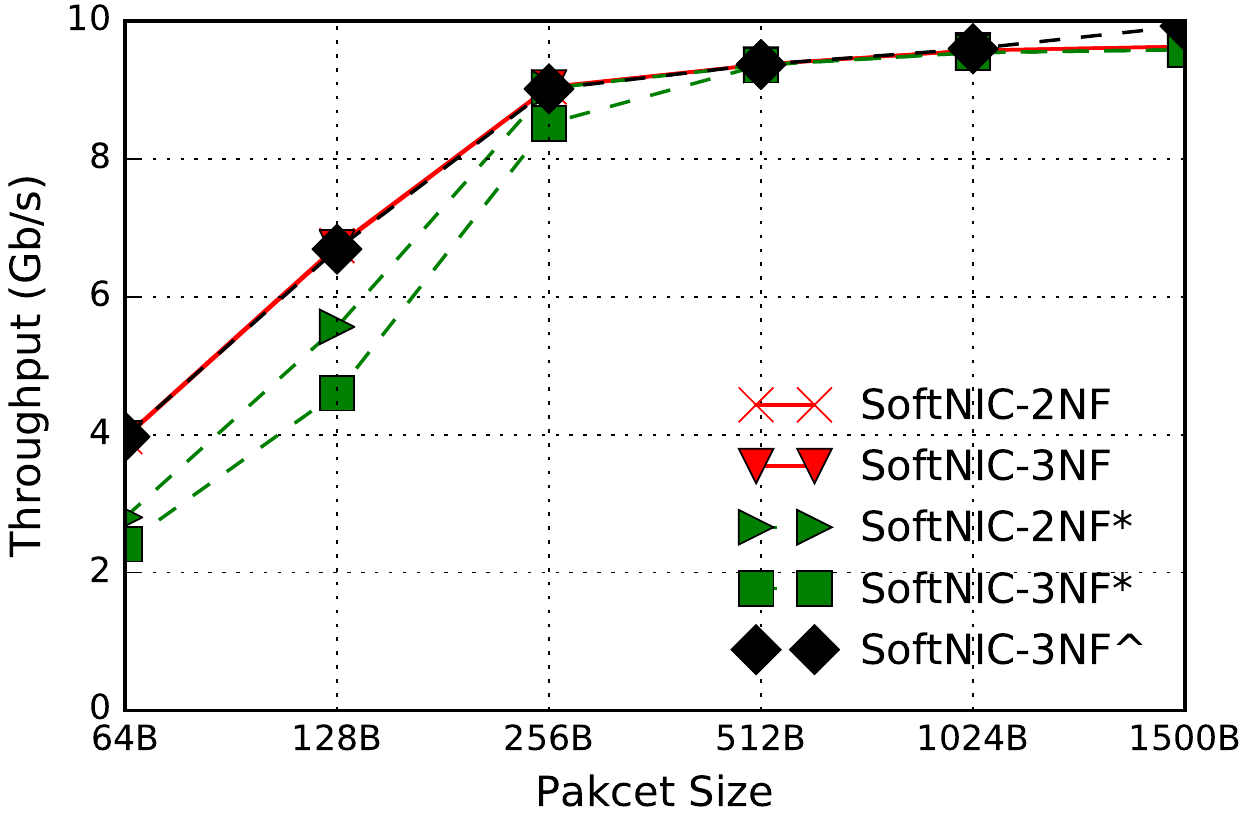}
    \caption{Throughput of NF chaining with the NUMA effect. ``*'' here denotes the case when the vCPUs belong to different sockets. ``\string^'' here denotes the case when NFs are chained up via NICs of different servers to avoid penalty from NUMA.}
    \label{fig:rst_chain2}
    \vspace{-1.5mm}
\end{figure}

We perform another measurement to evaluate the effect of NUMA on NF chaining. 
Figure~\ref{fig:rst_chain2} shows the result for SoftNIC as a case study. 
NUMA has a significant impact on performance. For the 2-NF chain, assigning two vCPUs and TCs to different sockets cuts the throughput of 64B packet by nearly half. For the 3-NF chain (the third VNF runs in a different socket than the first two), line rate is only reached for 512B and larger packets. The performance discrepancy is mainly because operations between different NUMA sockets can cause cache misses and ping pong effect \cite{levinthal2009performance}. \zhixiong{To mitigate NUMA effect, we attempt to bridge NFs in a chain via NICs across servers. For example, in a 3-NF chain, NF1 and NF2 are located on server A on the same NUMA socket, and NF3 is located in server B. The two servers are connected by 10GbE NIC. We observe that this eliminates the NUMA effect: throughput of the chain is identical to the case when all 3 NFs share the same CPU socket as shown in Figure~\ref{fig:rst_chain2}. }

\revise{To summarize, the results here show that SoftNIC works adequately with small performance drop in a NF chain, while ClickOS's throughput becomes lower with longer chains. They also demonstrate the importance of carefully assigning cores to VMs of the chain due to NUMA, which implies that it is not always best to colocate VNFs of a chain on the same server. A practical strategy is to place them on different servers to avoid NUMA effect. These observations are useful for real NFV deployment. }

\subsection{NF Colocation}
\label{sec:colocation}

We also measure the performance when multiple NFs colocate on the same server. This is another common deployment scenario of NFV. In the experiments here, we instantiate multiple VMs, bundling each of them to an independent packet generator in the same server, and measure the aggregated throughput.  
For SoftNIC, we build {\tt pkt-gen} in separate VMs and connect them to L3FWD VNFs by independent TCs. On the other hand for ClickOS, we directly use {\tt pkt-gen} to generate packets on the VALE switch connected to the VM. We pin {\tt pkt-gen} and the corresponding VM to the same CPU socket for better performance. Note this is the only scenario where we generate packets at the hypervisor. We scale to at most 3 bundles beyond which the number of cores on our CPU is not enough (our CPU has 16 cores: SoftNIC needs 4 cores for each bundle, and the hypervisor needs cores too).

\begin{figure}[ht]
    \centering
    \vspace{-2mm}
    \includegraphics[width=0.9\linewidth]{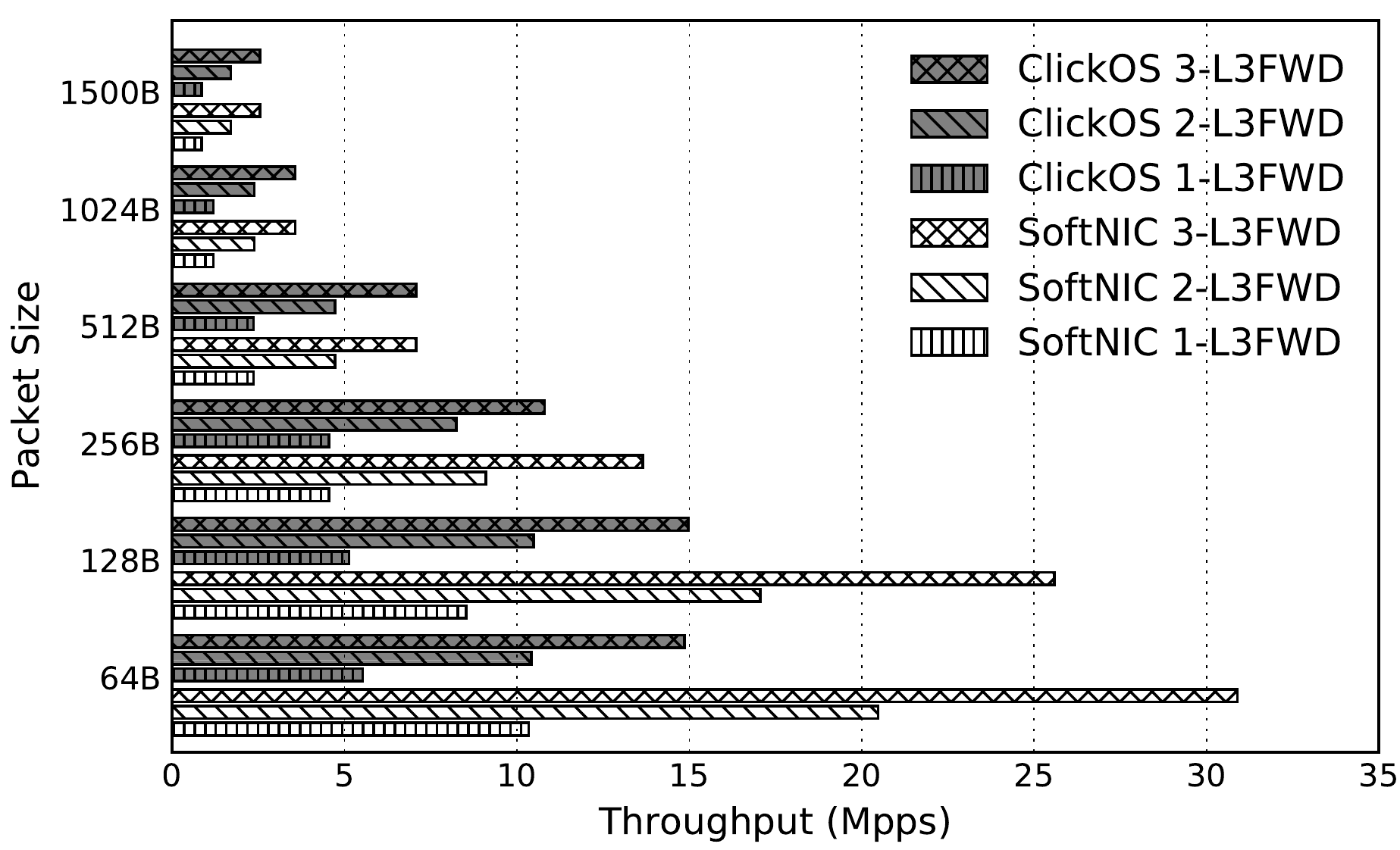}
    \caption{Aggregated throughput of multiple colocating L3FWD. }
    \label{fig:scaling}
    \vspace{-4mm}
\end{figure}

As we observe in Figure~\ref{fig:scaling}, both NFV dataplanes perform very well. In almost all results, throughput scales linearly as we colocate more bundles of VMs and packet generators. 
This demonstrates that current technologies provide satisfactory performance isolation and guarantee with multi-core CPUs for realizing NFV.


\section{Related Work}
\label{sec:related}




We introduce related work on NFV dataplane other than SoftNIC and ClickOS now. 
NetVM \cite{HRW14} is another NFV platform based on DPDK and KVM, similar to SoftNIC. It provides high-speed inter-VM communication with zero-copy through shared huge pages. We plan to study its performance when the code becomes available. Systems such as ptnetmap \cite{garzarella2015virtual} and mSwitch \cite{honda2015mswitch} based on netmap address efficient transfer between VMs in a single server. E2 \cite{PLHJ15} is a general NFV management framework focusing on NF placement, scheduling, scaling, etc. Its dataplane uses SoftNIC.  

Our measurement study provides performance comparison across solutions with actual VNFs and complements existing work that evaluates their own system with mostly L2 forwarding. There is little measurement study on NFV in general. Wu et al. design PerfSight \cite{WHA15} as a diagnostic tool for extracting comprehensive low-level information regarding packet processing performance of the various elements. It focuses on virtualization layer (KVM) without integrating with any NFV dataplane such as SoftNIC and ClickOS.

\section{Conclusion}

In this paper, we conducted a measurement study on the performance of SoftNIC and ClickOS. Both dataplanes are capable of achieving 10G line rate with medium and large packets, and scaling performance with multiple colocating VNFs. They have performance issues in the small packet regime and NF chaining scenario, which may become more severe in high speed networks. 
We proposed to fundamentally address the limitation by architecturing the software dataplane for horizontal performance scaling, in order to better utilize multiple cores and NIC queues.

Our study can be extended in many directions. 
One possibility is to consider more complex NFs such as NAT, VPN, etc., and more metrics such as processing delay. We also plan to further investigate the chaining scenario and identify ways to improve performance. 



\section{Acknowledgment}
We are very thankful to members of SoftNIC and ClickOS teams, especially to Sangjin Han from UC Berkeley, and Filipe Manco from NEC Laboratories Europe.
\bibliographystyle{abbrv}
\balance

\bibliography{IEEEabrv,main}

\end{document}